\documentclass{SciPost}

\hypersetup{
    colorlinks,
    linkcolor={red!50!black},
    citecolor={blue!50!black},
    urlcolor={blue!80!black}
}

\usepackage[bitstream-charter]{mathdesign}
\usepackage{tikz-cd}
\usepackage{xspace}
\usepackage{multicol}
\DeclareSymbolFont{usualmathcal}{OMS}{cmsy}{m}{n}
\DeclareSymbolFontAlphabet{\mathcal}{usualmathcal}
\newcommand{\Pu}{\ensuremath{^{241}\textrm{Pu}}\xspace}
\newcommand{\T}{\ensuremath{^{3}\textrm{H}}\xspace}

\newcommand{\Tm}{$^{171}\textrm{Tm}$~}
\newcommand{\Sm}{$^{151}\textrm{Sm}$~}

\fancypagestyle{SPstyle}{
\fancyhf{}
\lhead{\colorbox{scipostblue}{\bf \color{white} ~SciPost Physics }}
\rhead{{\bf \color{scipostdeepblue} ~Submission }}

\fancyfoot[C]{\textbf{\thepage}}
}

\begin{document}

\pagestyle{SPstyle}

\begin{center}{\Large \textbf{\color{scipostdeepblue}{
Measuring the Neutrino Mass with \texorpdfstring{$^{241}$Pu}{Pu-241}
}}}\end{center}

\begin{center}\textbf{
Mustafa Anaam\textsuperscript{1},
Nicolo de Groot\textsuperscript{1} and
Inge van Rens\textsuperscript{1}
}\end{center}

\begin{center}
{\bf 1} Radboud University, Nijmegen, The Netherlands
\end{center}

\section*{\color{scipostdeepblue}{Abstract}}
\textbf{\boldmath{%
Tritium has long been the isotope of choice for measurements of the absolute neutrino mass. Its low nuclear mass leads to substantial recoil energy, limiting measurements with both molecular tritium ($T_2$) and atomic tritium bound to a substrate. With a beta-decay endpoint energy and half-life similar to those of tritium, but a much larger nuclear mass, $^{241}$Pu may offer an interesting alternative. Its principal challenge is background from the beta decay of $^{237}$U. We discuss strategies for suppressing this background and compare the projected neutrino-mass sensitivities of $^{241}$Pu and tritium as a function of the background suppression.
}}

\vspace{0.5\baselineskip}
\noindent\textbf{Keywords:} neutrino physics, neutrino mass, beta decay, tritium, plutonium

\vspace{\baselineskip}

\noindent\textcolor{white!90!black}{%
\fbox{\parbox{0.975\linewidth}{%
\textcolor{white!40!black}{\begin{tabular}{lr}%
  \begin{minipage}{0.6\textwidth}%
    {\small Copyright attribution to authors. \newline
    This work is a submission to SciPost Physics. \newline
    License information to appear upon publication. \newline
    Publication information to appear upon publication.}
  \end{minipage} & \begin{minipage}{0.4\textwidth}
    {\small Received Date \newline Accepted Date \newline Published Date}%
  \end{minipage}
\end{tabular}}
}}
}


\section{Introduction}
\label{sec:intro}

For decades tritium has been the preferred nuclide to measure the absolute neutrino mass. Its low endpoint energy of
18.6 keV, the relatively large fraction of events decaying near the kinematic endpoint, its simple atomic structure and
its favorable half-life of 12.3 years have all contributed to this fact. The current state-of-the-art measurement is done
by the KATRIN experiment and uses molecular tritium to set the current world's best direct neutrino-mass limit, $m_\nu<0.45$ eV at 90\% confidence level~\cite{katrin2025}. This measurement is still dominated by statistical uncertainty, but with more data it will be limited by the energy spread of the vibrational and rotational excitation spectrum of the daughter molecule $(^3\textrm{H}^3\textrm{He})^+$. The decaying tritium nucleus recoils against the electron and picks up 3.4 eV kinetic energy due to its relatively low mass. The recoil is enough to bring the daughter molecule into one of several excited states and leads to an energy spread of 0.36 eV.
This is why future experiments, like Project8~\cite{Project8-2017} and QTNM~\cite{qntm} are looking at atomic tritium.

The PTOLEMY experiment~\cite{Ptolemy2019} plans to use atomic \T loaded on graphene to measure the
mass and, at a later stage, observe the CNB. A study~\cite{Cheipesh2021} found that the zero-point motion of the \T bound
to the surface of graphene, or for that matter any surface, leads to an energy spread of $\Delta E \sim 0.5$ eV.
More precise calculations on the binding of tritium to graphene~\cite{Tozzini26} using density functional theory indicate that
the uncertainty in the energy can be evaluated and are not limiting for the measurement of the neutrino mass, but also brought up
additional sources of uncertainty from the ordering of the tritium atoms in case of less than complete coverage to the time
variation of the substrate as tritium atoms decay.

Alternative nuclides for the mass measurement should first of all have moderate beta-decay energy. For isotopes with
a decay energy of 10-100 keV, already an experimental precision of 1ppm is required.
A lifetime of 1-20 years is optimal since it leads to a reasonable event rate and relatively stable measurement
conditions. Rhenium-187 has the lowest energy of the known beta-decaying isotopes at 2.6 keV, but due to its lifetime of 60 Gyr, there would be no events close to the endpoint within a human lifetime.
In addition, the daughter isotope needs to be stable on the timescale of the experiment. Ruthenium-106
for example, does have a decay energy of 39.4 keV, and a half-life of 1.48 years,
but the daughter isotope Rhodium-106 decays within 30 seconds with a decay energy of several MeV and the
emission of several gammas, making a clean measurement of the endpoint impossible.  In ~\cite{Cheipesh2021} the investigation of nuclei with $A>100$ in particular \Tm and \Sm as candidate isotopes is proposed. One drawback here is the substantial higher energies
of the beta decay. The number of events in the endpoint of the spectrum scales with the inverse third power of the
total decay energy. This means \Tm and \Sm decays are suppressed by two orders of magnitude. Even if the experimental set-up could
be made to work for electrons of a much higher energy, it still would require a relative precision in the energy measurement
that is 4-5 times better than for \T.

\Pu was suggested as a more suitable alternative ~\cite{ndg-pu241} as it is much more similar to \T.
We investigate in this paper the potential of \Pu as a source isotope for the measurement of the absolute neutrino mass
and compare it with tritium. In the next session we discuss the beta decay of \Pu. In section~\ref{sec:background-suppression}
we discuss possible ways to reduce the background and in section~\ref{sec:sensitivity-study} we present a sensitivity study for
\Pu and \T. Section~\ref{sec:results} is devoted to the results, while section~\ref{sec:conclusion} summarizes our conclusions.

\section{Beta decay of tritium and \texorpdfstring{$^{241}$Pu}{Pu-241}}
\label{sec:beta-decay}
Plutonium-241 is an isotope that is created by double neutron capture on Plutonium-239.
About 12\% of the plutonium in spent nuclear fuel is \Pu.
It mainly undergoes beta decay to $^{241}\textrm{Am}$ with a decay energy of $20.78 \pm 0.17$ keV~\cite{audi1995} and a
half-life of $14.33 \pm 0.02$ years~\cite{DEBIEVRE1983111}, numbers very similar to those of \T.
What makes \Pu interesting for the mass measurement is the large
mass of the isotope and as a consequence its low recoil of 47 meV.

The daughter nuclide $^{241}\textrm{Am}$ decays trough $\alpha$ decay to $^{237}$Np with a half-life
of 432 years. $^{237}$Np has a half-life of 2 million years and can be considered stable.  In
addition, \Pu has a  small ($2.4 \times 10^{-5}$) probability to undergo $\alpha$ decay to $^{237}$U.
The energies of the $\alpha$ particles, and the subsequent $\gamma$'s, are well outside the relevant energy
window around the endpoint.  The biggest challenge comes from the $^{237}$U. It has several $\beta$
decays to $^{237}$Np with a half-life of 6.7 days and an energy up to 459 keV.  The total rate from
 the $^{237}$U is about five orders of magnitude smaller than for \Pu, but near the endpoint the \Pu
 spectrum is falling rapidly, while $^{237}$U is still on a plateau.  Around 20 eV before the endpoint,
 the rates become similar, and a mass measurement becomes impossible without special measures
 to remove the $^{237}$U on a timescale much shorter than its lifetime.

Given the similar decay energy and half-life, one would  expect the experimental setup for
atomic \T to work for \Pu as well. For instance the Project8 experiment plans to reach its ultimate
precision using an atomic \T trap and will have to balance statistics, which favors a high density,
against the background of recombination of tritium into T$_2$. \Pu, in the form of gaseous PuF$_6$
would be an alternative where no such compromise is needed. For the 
QTNM experiment, a beamline with \Pu atoms or ions is an option.
For PTOLEMY, the design of the RF system, electromagnetic filter and the calorimeter or hemispherical
analyzer would be identical. One  requirement is  to have a precise reference voltage between the target and the exit of the electromagnetic filter. This in fact is easier using a metallic target like plutonium than for graphene.

\begin{figure}[t]
  \centering
  \includegraphics[width=0.8\linewidth]{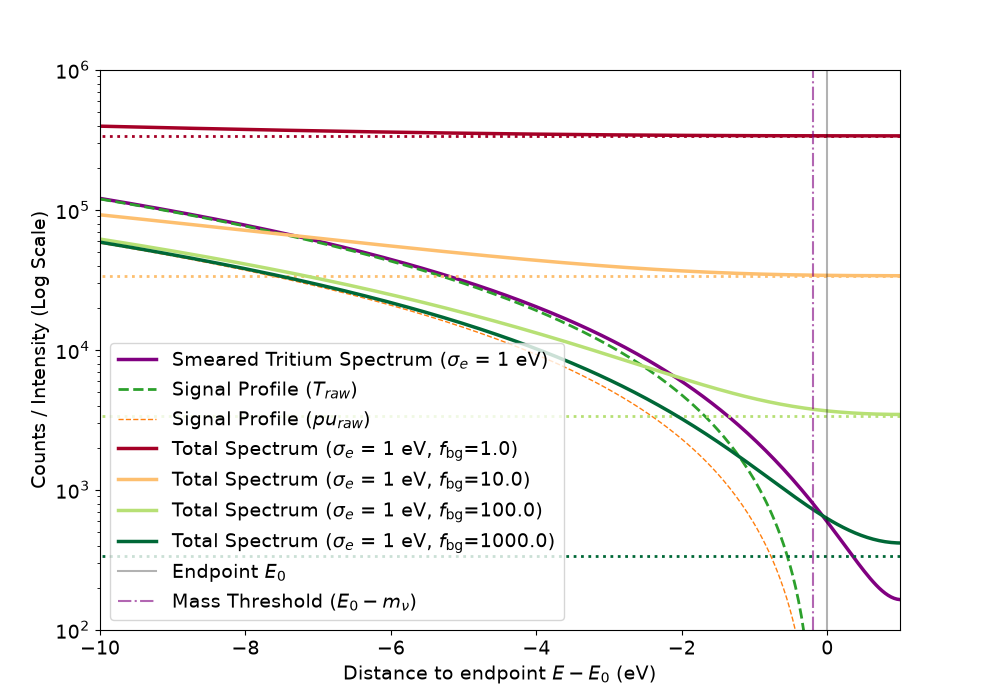}
  \caption{Spectral decomposition near the endpoint for $m_\nu=0.2\,\mathrm{eV}$, with the plutonium source normalized by its total activity. Several levels of background suppression are shown.}
  \label{fig:spectrum}
\end{figure}

\section{Background suppression}
\label{sec:background-suppression}
\begin{figure}[t]
  \centering
  \begin{minipage}[t]{0.48\linewidth}
    \centering
        \includegraphics[width=\linewidth]{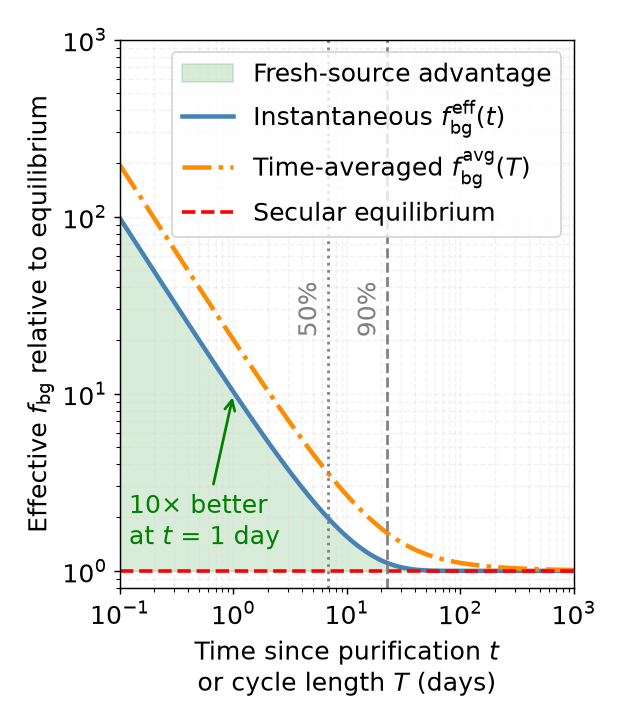}
  \\ (a)
  \end{minipage}\hfill
  \begin{minipage}[t]{0.48\linewidth}
    \centering
     \includegraphics[width=\linewidth]
    {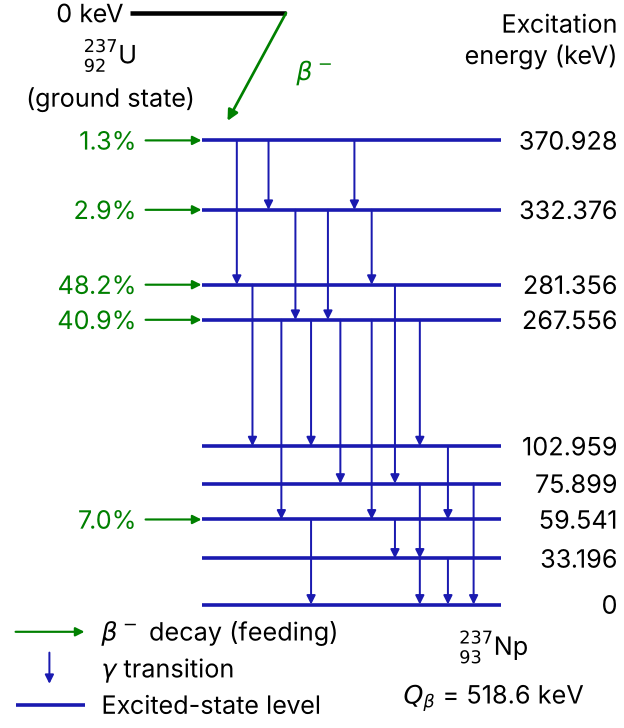}
  \\ (b)
  \end{minipage}
  \caption{
\textbf{(a):} Effective background suppression relative to secular
equilibrium for a freshly purified $^{241}$Pu source. The solid curve is the
instantaneous suppression $f_{\rm bg}^{\rm eff}(t)$ at time $t$ after
purification. The dash-dotted curve is the suppression $f_{\rm bg}^{\rm avg}(T)$
averaged over a cleaning cycle of length $T$. The vertical lines mark when the
$^{237}$U population reaches 50\% and 90\% of its equilibrium value, and the
shaded region indicates the fresh-source advantage.
 \textbf{(b):} Simplified decay scheme of $^{237}$U to $^{237}$Np, the
$\beta^-$-decay feeding fractions are shown in green and the subsequent
$\gamma$-ray transitions in blue~\cite{np237data}.
}
\label{fig:background}
\end{figure}

The dominant intrinsic background for a \Pu endpoint measurement is not the direct $\alpha$ activity itself, but the
$^{237}$U that is continuously produced in the rare $\alpha$ branch of \Pu.  We denote the branching fraction by
$b \simeq 2.5\times 10^{-5}$.  Since the half-life of $^{237}$U is only 6.75 days, corresponding to a mean lifetime
$\tau_U \simeq 9.74$ days, its population is set by a competition between production from the parent source and
removal by its own decay.
The decay scheme for the subsequent decays is illustrated in  Fig.~\ref{fig:background}b.
In the absence of active mitigation,
\begin{equation}
  \frac{dN_U}{dt} = b\lambda_{\rm Pu}N_{\rm Pu} - \lambda_U N_U,
\end{equation}
and the activity builds up on a timescale of order $\tau_U$ towards the equilibrium value.  This is small compared
with the total \Pu activity, but it is problematic near the endpoint because the \Pu spectrum falls rapidly while the
higher-energy $^{237}$U beta spectrum remains slowly varying across the same region.  A useful suppression strategy
therefore has two goals: it should reduce the number of $^{237}$U atoms in the source, and it should make the residual
component distinguishable from the neutrino-mass signal.

A simple approach is to periodically clean or refresh the source.  If the uranium is removed after a fixed cycle time
$T$, the $^{237}$U population repeatedly starts near zero and grows as $1-\exp(-t/\tau_U)$.  Averaged over a cycle, the
remaining fraction of the equilibrium population is
\begin{equation}
  \frac{\langle N_U\rangle}{N_U^{\rm eq}} =
  1 - \frac{1 - e^{-T/\tau_U}}{T/\tau_U} .
  \label{eq:cycle_average}
\end{equation}
For a two-day cleaning cycle this is about 0.11.  Thus the raw background associated with the $\alpha$ branch is reduced
from roughly $2.5\times 10^{-5}$ to the few-$10^{-6}$ level before making use of any spectral or timing information. The time dependence of the suppression is shown in {Fig.~\ref{fig:background}a}, relative to secular equilibrium, for a
freshly purified $^{241}$Pu source. Immediately after purification the
$^{237}$U content is negligible, so the instantaneous suppression
$f_{\rm bg}^{\rm eff}(t)$ (solid curve) is large: about a factor of ten after
one day. It decreases as $^{237}$U regrows, and the population reaches 90\%
of its equilibrium value after about 22 days. The dashed curve shows the
suppression $f_{\rm bg}^{\rm avg}(T)$ averaged over a cleaning cycle of length
$T$, given by the inverse of Eq.~(\ref{eq:cycle_average}). It lies above the
solid curve because each cycle starts almost free of uranium; for a two-day
cycle it is about ten, compared with 5.4 at the end of the cycle. The averaged
value is the effective $f_{\rm bg}$ that periodic source refreshing provides
to a time-integrated analysis.

A stronger mitigation strategy is continuous extraction.  This can be described by adding an effective removal rate
$\lambda_{\rm rem}$,
\begin{equation}
  \frac{dN_U}{dt} = b\lambda_{\rm Pu}N_{\rm Pu} - (\lambda_U+\lambda_{\rm rem})N_U .
\end{equation}
The steady-state population is then suppressed relative to the no-removal case by
\begin{equation}
  \frac{N_U^{\rm cont}}{N_U^{\rm eq}} = \frac{\lambda_U}{\lambda_U+\lambda_{\rm rem}}
  = \frac{\tau_{\rm rem}}{\tau_U+\tau_{\rm rem}},
\end{equation}
where $\tau_{\rm rem}=1/\lambda_{\rm rem}$.  Removal on a one-day timescale leaves about 9\% of the equilibrium uranium
population, while removal on a 0.1-day timescale leaves about 1\%.  Starting from the $2.5\times10^{-5}$ branching
fraction, this corresponds to an effective steady-state level of order $10^{-7}$--$10^{-6}$, depending on the engineered
extraction rate.

 In several existing or proposed neutrino-mass experiments, the source geometry naturally provides opportunities for active suppression of daughter nuclei.
 For a KATRIN-like spectrometer and for Project8 one could consider a gaseous source based on a volatile plutonium compound such as PuF$_6$, the alpha-decay branch producing $^{237}$U may provide a natural mechanism for daughter removal. The approximately 100~keV recoil energy of the $^{237}$U nucleus greatly exceeds molecular binding energies and is expected to fragment the parent molecule while simultaneously producing highly ionized daughter species. As a result, a significant fraction of the newly created uranium atoms may appear as charged ions or ionized molecular fragments. Continuous gas circulation through an electrostatic collection region could therefore remove $^{237}$U on timescales of seconds to minutes, substantially shorter than the 6.75-day half-life of the daughter isotope.

 A PTOLEMY-like approach employing a liquid or flowing source could be circulated through a radiochemical purification system, for example using ion-exchange, solvent-extraction, or electrochemical techniques developed in the nuclear fuel cycle. Since plutonium and uranium exhibit different oxidation-state chemistry, the daughter nuclei could be continuously removed from the active measurement volume. The achievable suppression would be determined by the circulation time of the loop, which could potentially be reduced to hours or less in a suitably engineered system. Alternatively, in a solid source made of a thin layer (up to 10 nm) the $\alpha$ recoil would eject a significant fraction of the $^{237}$U from the source after which it can be removed by sputtering, reactive chemistry, or surface getters.

 The QTNM concept may provide perhaps the most direct method for daughter rejection.  With a beamline as a source, the large recoil energy associated with the alpha-decay branch will eject newly formed $^{237}$U atoms. Additional electrostatic or magnetic cleaning stages within the beamline could intercept recoil-produced ions and transport them to dedicated catcher surfaces. In such a configuration the effective removal time would be set by the transport time through the apparatus rather than by the nuclear half-life, potentially suppressing the $^{237}$U population by several orders of magnitude relative to secular equilibrium.

The time dependence also provides a handle in the statistical analysis.  Each event can be labeled not only by its
energy but also by the time elapsed since the last cleaning step.  The expected rate may then be written schematically as
\begin{equation}
  \lambda(E,t) = A_{\rm Pu} S(E;m_\nu^2) + A_U^{(0)}\left(1-e^{-t/\tau_U}\right)B(E),
\end{equation}
where $S(E;m_\nu^2)$ is the endpoint spectrum of \Pu and $B(E)$ is the local shape of the $^{237}$U background.  The
important point is that the signal is constant in time on the scale of a cleaning cycle, whereas the uranium contribution
has a predictable growth curve.  A time-stamped likelihood therefore does more than subtract a normalization: it reduces
the degeneracy between a distortion of the endpoint spectrum and a residual $^{237}$U component. Such a background reduction method could be applied in experiments relying on a solid target, with tritium bound to the substrate like PTOLEMY.

An additional event-by-event discriminator is provided by the characteristic $\gamma$ rays emitted when $^{237}$U
$\beta$ decay populates excited states of $^{237}$Np, as illustrated in {Fig.~\ref{fig:background}b}.
A surrounding scintillator or semiconductor
detector could identify these photons in prompt coincidence with an electron in the endpoint detector, allowing the
corresponding event to be vetoed.  The desired $^{241}$Pu $\beta$-decay signal does not produce the same $^{237}$Np
de-excitation cascade, so this coincidence requirement can suppress the uranium background without relying on its
electron-energy spectrum.  The coincidence window must be chosen
to retain the prompt cascade while keeping accidental vetoes and the associated loss of signal exposure acceptably small. While all $^{237}$U decays are to an exited state of $^{237}$Np, only 54\% produce one or more $\gamma$ rays~\cite{np237data}. This is caused by the the large probability for internal conversion (IC) during the de-excitation of the daughter nucleus $^{237}$Np. Since every $\beta$ decay populates an excited state of $^{237}$Np, the subsequent electromagnetic cascade proceeds through a combination of $\gamma$ emission and internal conversion. In contrast to the continuous $\beta$ spectrum, conversion electrons are emitted with well-defined energies,

\begin{equation}
E_{\mathrm{CE}} = E_{\mathrm{tr}} - E_{\mathrm{bind}}(n\ell),
\end{equation}

where $E_{\mathrm{tr}}$ is the nuclear transition energy and $E_{\mathrm{bind}}$ is the binding energy of the atomic shell involved in the conversion process.

Using the evaluated decay scheme of Ref.~\cite{np237data}, we reconstructed the complete electromagnetic cascade and calculated the probability that at least one $\gamma$ ray or conversion electron is emitted following the $\beta$ decay. The calculation is performed recursively over the decay cascade, using the measured transition intensities and internal conversion coefficients.

The resulting ideal emission probabilities are

\begin{align}
P(\gamma) &\approx 54\%,\\
P(\mathrm{CE}) &\approx 93\%.
\end{align}

The probability for observing either a $\gamma$ ray or a conversion electron is therefore essentially unity in the ideal case. Internal conversion is thus considerably more powerful than a $\gamma$-only veto.

Several conversion-electron lines are particularly intense. The dominant example is the
208~keV transition from the 267.6~keV level to the 59.5~keV level, which has a total internal conversion coefficient $\alpha_T\simeq2.98$. This transition produces a strong K-shell conversion line at approximately 89.3~keV with an intensity of about 50 electrons per 100 $^{237}$U decays. Additional prominent groups occur around 38--42~keV, 54--59~keV and 186--208~keV.

Table~\ref{tab:ICveto} summarizes the calculated probabilities for different conversion-electron thresholds. These values assume perfect detection efficiency for emitted particles and therefore represent the maximum achievable veto efficiency.

\begin{table}[ht]
\centering
\caption{Ideal probabilities for detecting at least one conversion electron above a given threshold following a $^{237}$U decay.}
\label{tab:ICveto}
\begin{tabular}{cc}
\hline
Threshold & Tagged events \\
$E_{\rm CE}$ & ($\gamma$ or CE) \\
\hline
0 keV   & 100\% \\
5 keV   & $\approx100\%$ \\
10 keV  & $\approx100\%$ \\
20 keV  & 99\% \\
50 keV  & 96\% \\
100 keV & 65\% \\
\hline
\end{tabular}
\end{table}

Experimentally, the achievable veto efficiency will be reduced by the finite escape probability of conversion electrons from the radioactive source, detector acceptance and intrinsic detection efficiency. These effects are expected to depend strongly on the source thickness. Nevertheless, the large intrinsic probability for internal conversion suggests that a coincidence veto based on prompt conversion electrons, possibly supplemented by characteristic Np X-rays and $\gamma$ rays, could suppress the $^{237}$U background substantially more efficiently than a conventional $\gamma$ veto alone.

In the sensitivity estimates below we therefore treat the $^{237}$U contribution as a background whose normalization can
be reduced by an explicit suppression factor and, in realistic operation, further constrained by its time dependence.
The relevant question is not whether the $^{237}$U branch exists, but what suppression factor is required before the
remaining background no longer dominates the neutrino-mass sensitivity.

\section{Sensitivity Study}
\label{sec:sensitivity-study}

\subsection*{Spectral Model}

The sensitivity to the effective neutrino mass was evaluated using simulated endpoint spectra of
$^{241}$Pu and tritium. The analysis was performed in an energy window extending from
$E_0-40~\mathrm{eV}$ to the endpoint energy $E_0=21.78~\mathrm{keV}$, sampled with 1000 uniformly spaced bins.

For a given neutrino mass hypothesis $m_\nu$, the unsmeared differential signal spectrum was modeled as

\begin{equation}
S(E;m_\nu)
=
N_{\mathrm{iso}}\,C_{\mathrm{iso}}\,
\sqrt{\max\!\left[(E_0-E)^2-m_\nu^2,\,0\right]}
\,
\max(E_0-E,0),
\end{equation}

where $N_{\mathrm{iso}}$ is the total number of source decays contributing during the measurement
period and $C_{\mathrm{iso}}$ is an isotope-dependent normalization constant.
We use $C_{\mathrm{Pu}}=3.9\times10^{-13}$ for plutonium and
$C_{\mathrm{T}}=8.6\times10^{-13}$ for tritium. These values were determined using BetaShape software\cite{Mougeot-betashape}.

For the $^{241}$Pu source, the total number of decays over the data-taking period was estimated as

\begin{equation}
N_{\mathrm{Pu}}
=
\frac{m_{\mathrm{Pu}}N_A}{241}
\,
\frac{t_{\mathrm{run}}}{\tau_{\mathrm{Pu}}},
\end{equation}

where $m_{\mathrm{Pu}}=100~\mu\mathrm{g}$ is the source mass,
$N_A$ is Avogadro's number,
$t_{\mathrm{run}}=3$ years is the exposure,
and $\tau_{\mathrm{Pu}}=20.68$ years is the lifetime associated with the decay process.

\subsection*{Background Model}

For the $^{241}$Pu analysis, a flat component represented the $^{237}$U background, with spectral density $\rho_U$ and normalization

\begin{equation}
A_U=\frac{N_{\mathrm{Pu}}f_\alpha}{f_{\mathrm{bg}}},
\end{equation}

where
$f_\alpha = 2.5\times10^{-5}$ is the assumed branching fraction,
$\rho_U = 9.1\times10^{-6}$ is the background spectral density,
and $f_{\mathrm{bg}}$ is a user-defined scaling factor controlling the overall background level.
No background contribution was included in the tritium case.

The total unsmeared spectrum is therefore $N(E;m_\nu)=S(E;m_\nu)+B(E)$.

\subsection*{Detector Energy Resolution}

Finite detector resolution was modeled through convolution of the spectrum with a Gaussian response function. In practice, the spectrum was smoothed using a Gaussian kernel with width $\sigma_E$, resulting in the reconstructed spectrum $N_{\mathrm{rec}}(E;m_\nu,\sigma_E)$. Sensitivity studies were carried out for energy resolutions  between 0.05 and 10 eV, spaced logathmically.

\subsection*{Asimov Data Sets}

Expected sensitivities were evaluated using the Asimov data set formalism. For each assumed true neutrino mass
$m_\nu^{\mathrm{true}}$, a data set was generated directly from the model,

\begin{equation}
N_i^{\mathrm{A}}
=
N_{\mathrm{rec}}
(E_i;
m_\nu^{\mathrm{true}},
\sigma_E).
\end{equation}

The statistical uncertainty assigned to each bin was

\begin{equation}
\sigma_i
=
\sqrt{
N_i^{\mathrm{A}}
+
\beta B_i
},
\end{equation}

where a background-normalization factor $\beta=0.01$ was included to account for uncertainty in the background normalization. Empty bins were assigned a unit uncertainty to avoid numerical singularities.

\subsection*{Chi-Squared Statistic}

The agreement between a neutrino-mass hypothesis and the Asimov spectrum was quantified using

\begin{equation}
\chi^2(m_\nu)
=
\sum_i
\left[
\frac{
N_i^{\mathrm{A}}
-
N_i^{\mathrm{model}}(m_\nu)
}{
\sigma_i
}
\right]^2 ,
\end{equation}

where
$N_i^{\mathrm{model}}(m_\nu)$
is the reconstructed spectrum corresponding to the tested mass hypothesis.

\subsection*{Expected 90\% C.L. Upper Limits}

The expected neutrino-mass upper limit was obtained using a profile-likelihood approximation based on the Asimov data set.
Since the Asimov spectrum is generated at the true parameter value,
the minimum chi-squared occurs at $\chi^2_{\mathrm{min}}=\chi^2(m_\nu^{\mathrm{true}})$.

The one-sided 90\% confidence limit was defined by the condition

\begin{equation}
\Delta\chi^2
=
\chi^2(m_\nu^{90})
-
\chi^2_{\mathrm{min}}
=
2.71,
\end{equation}

which corresponds to the standard one-parameter 90\% confidence criterion.
The upper limit $m_\nu^{90}$ was determined by scanning upward from the assumed true mass until the threshold was reached.

\subsection*{Discovery Sensitivity}

The discovery sensitivity was evaluated by testing the massless-neutrino hypothesis against an Asimov data set generated with a non-zero neutrino mass.
We used the test statistic $q_0=\chi^2(m_\nu=0)$. Assuming Wilks-type asymptotic behavior, the corresponding Gaussian-equivalent significance was $Z=\sqrt{q_0}$.

A detection was considered achievable when $Z \ge 3$,
corresponding to a $3\sigma$ discovery threshold.



\section{Results}
\label{sec:results}

We investigate three main research questions:
\begin{enumerate}
  \item How do \T and \Pu compare for the same source activity?
  \item How do \T and \Pu compare for the same source activity close to the endpoint?
  \item How much extra \Pu is needed to be competitive with \T?
\end{enumerate}

As can be seen in Figure~\ref{fig:spectrum} the activity for \T is near the endpoint is more than twice as high as for \Pu for the same overal activity. This has two reasons. First of all the endpoint energy for \Pu is 17\% higher that for \T and the spectrum near the endpoint scales with $1/E_0^3$. Secondly the coulomb interaction is much more significant for \Pu and that pulls the spectrum to lower values.

\begin{figure}[ht]
  \centering
  \includegraphics[width=\linewidth]{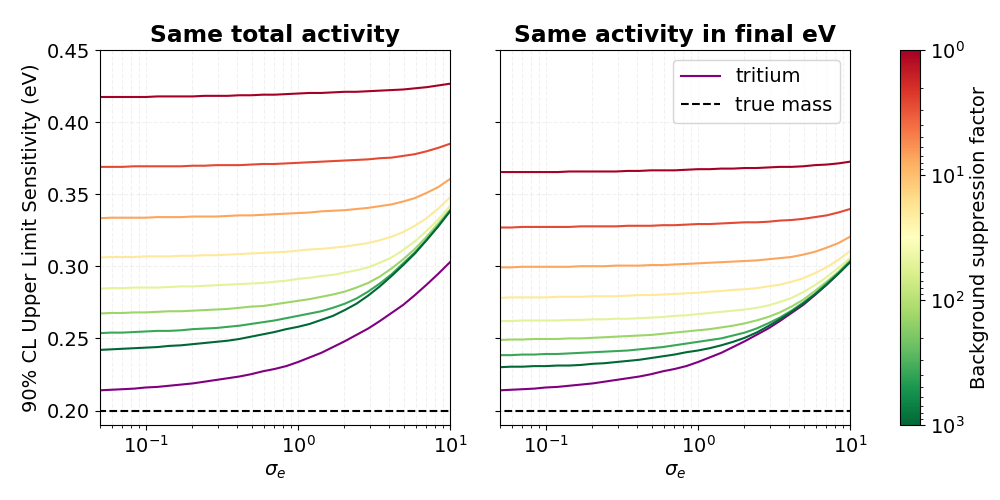}
  \caption{Expected 90\% confidence-level upper-limit sensitivity for \Pu and \T if the true neutrino mass is 0.2 eV. Higher background suppression improves the neutrino mass sensitivity. \textbf{Left:} For equal total activity. \textbf{Right:} For equal activity in the endpoint.}
  \label{fig:sensitivity-comparison}
\end{figure}

The sensitivity comparison for the same total activity is shown in the left panel of Fig.~\ref{fig:sensitivity-comparison}. Tritium has a far better neutrino mass sensitivity than plutonium which is the result 
of the statistical effect and background for \Pu while tritium has none. Even if we were able to perfectly suppress the background of the plutonium decay, the tritium decay would still have a higher event rate near the endpoint and thus a higher sensitivity to the neutrino mass. We see that the effect becomes less significant for larger experimental errors when the relative effect of the background is diluted.

If instead we scale the amount of plutonium we compare to tritium with the activity in the final eV instead of the total activity, we can remove some of the disadvantage of plutonium compared to tritium. Now, the total Pu+U spectra converge to the tritium spectrum for increasing background suppression.  The neutrino mass sensitivity comparison in this case can be found in the right plot of Fig.~\ref{fig:sensitivity-comparison}. For sufficiently high background suppression, we can see that the sensitivity for tritium and plutonium of a low resolution detector converge as the effects of smearing due to low resolution become dominant over the effects of the uranium background. For more sensitive detectors, we see that tritium keeps its advantage.

\begin{figure}[ht]
  \centering
  \includegraphics[width=\linewidth]{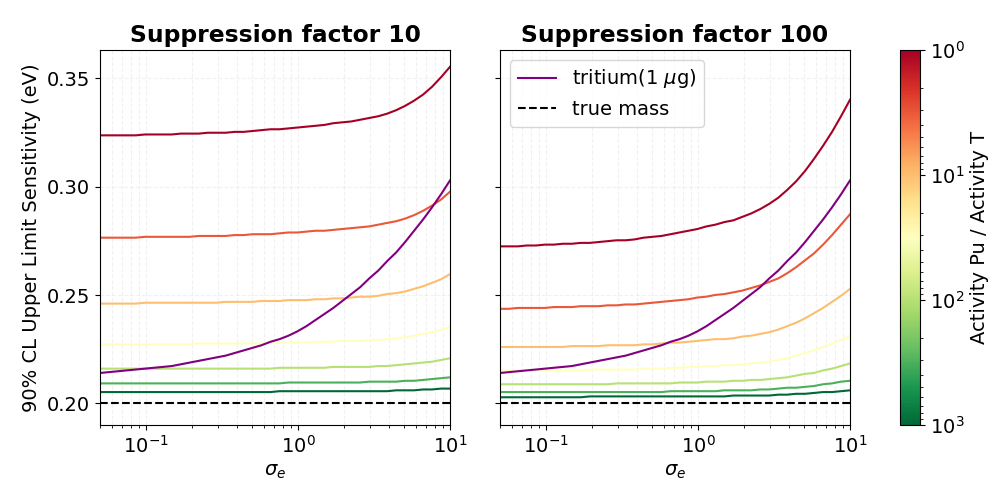}
  \caption{Expected 90\% confidence-level upper-limit sensitivity for \Pu and \T if the true neutrino mass is 0.2 eV. Depending on the detector resolution we need to increase the activity of \Pu by a certain amount to reach a higher sensitivity than \T. This comparison is shown for f=10 and f=100.}
  \label{fig:mass-comparison}
\end{figure}

If we use more target material, resulting in a higher activity, our sensitivity to the neutrino mass will go up. There might be some experiment designs where scaling the amount of \Pu than the amount of \T in the experiment. Therefore, we want to know how much more plutonium we need in order to reach the same sensitivity as tritium. How much more activity do we need to compensate for the background? The additional activity required for plutonium to match tritium sensitivity is shown in Fig.~\ref{fig:mass-comparison} for background suppression 10 and 100 . If we increase the amount of plutonium we use, we improve the sensitivity, and at some point become more sensitive than the reference tritium source. How much plutonium we need for this depends on the detector resolution and the uranium background suppression.

\begin{figure}[h!]
\centering
\includegraphics[width=\textwidth]{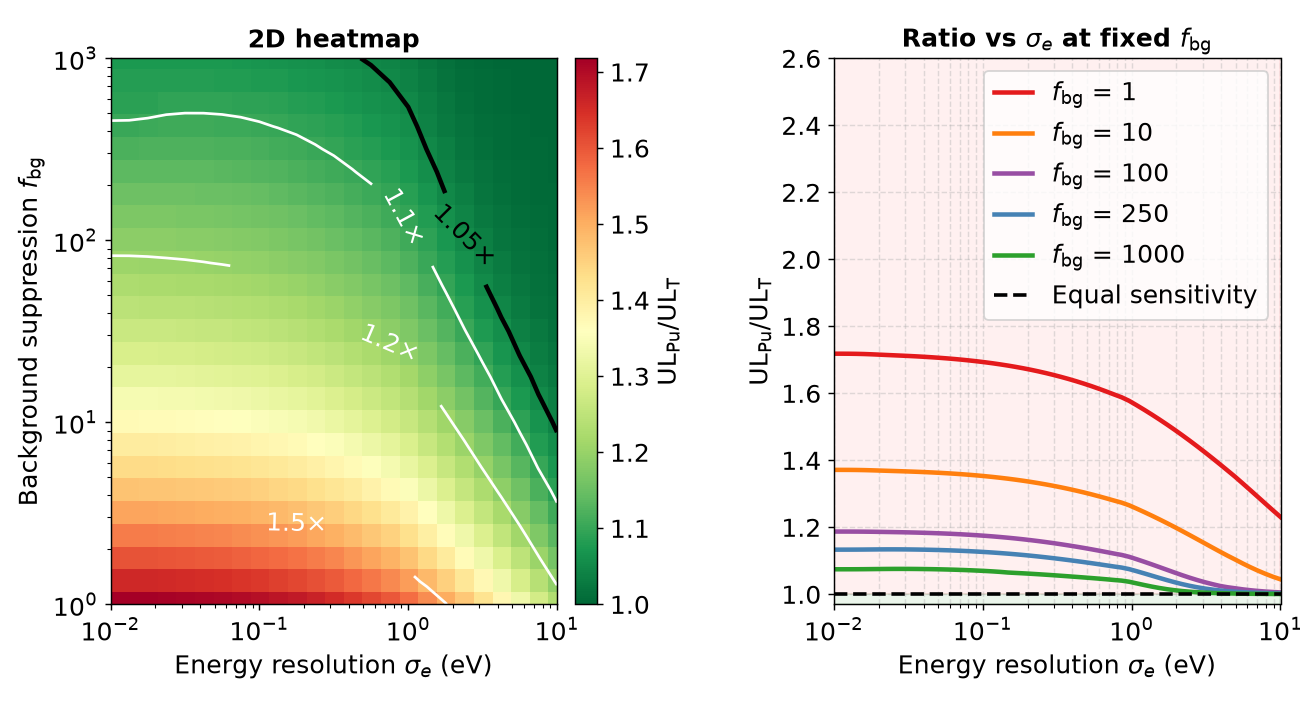}
\caption{Sensitivity ratio $\mathrm{UL_{Pu}} / \mathrm{UL_T}$ comparing
$^{241}$Pu and tritium under the endpoint activity normalization
($N_{\rm Pu} C_{\rm Pu} = N_T C_T$, corresponding to 206~$\mu$g of
$^{241}$Pu vs 1~$\mu$g of tritium), $m_{\nu}^{\mathrm{true}} = 0.2\text{ eV}$, 3-year exposure. \textbf{Left:} Two-dimensional heatmap as a function of energy resolution $\sigma_e$ and background suppression factor $f_{\mathrm{bg}}$, with contours marking ratio thresholds of 1.05, 1.1, 1.2, and 1.5. \textbf{Right:} Sensitivity ratio as a function of $\sigma_e$ for five fixed background suppression factors $f_{\mathrm{bg}}$, where the dashed horizontal line indicates equal sensitivity.}
    \label{fig:sensitivity_ratio_endpoint_mnu02}
\end{figure}

  The sensitivity ratio $\mathrm{UL}_{\rm Pu}/\mathrm{UL}_{\rm T}$ under endpoint normalization ($N_{\rm Pu}C_{\rm Pu}=N_{\rm T}C_{\rm T}$), corresponding to 206~$\mu$g of $^{241}$Pu, is shown in Fig.~\ref{fig:sensitivity_ratio_endpoint_mnu02}. The ratio remains above unity across the entire parameter space, indicating that tritium retains better sensitivity than $^{241}$Pu, even when both sources are normalized to produce the same spectral amplitude near the endpoint.

As shown in the right panel, the ratio decreases with increasing $\sigma_e$ and $f_{\rm bg}$. For the strongest background suppression ($f_{\rm bg}=1000$) and the largest energy-resolution width ($\sigma_e\sim10$~eV), the ratio approaches unity, indicating that $^{241}$Pu comes closest to matching the sensitivity of tritium under this normalization.

\section{Conclusion}
\label{sec:conclusion}
Tritium is still the most promising isotope for a neutrino mass measurement, as it has been since the first neutrino mass measurement experiments in 1948. The combination of the low decay energy, suitable lifetime, lack of secondary decays and relatively high availability of the isotope makes it hard to beat. In this work, however, we have probed the conditions under which \Pu could be a competitive alternative. \Pu has a similar decay energy and lifetime to \T, but is hampered by its rare alpha decay to $^{237}$U. In a setup where the $^{237}$U could easily and reliably be removed after the alpha decay, \Pu can be an alternative for \T. If additionally the activity of \Pu in this setup can be scaled up more easily than the mass of \T, it might even have a better sensitivity to the neutrino mass, depending on the activity scaling and the detector resolution.

In order for \Pu to be competitive to \T in its exclusion limits on the neutrino mass, the $^{237}$U background need to be suppressed by at least a factor 10, although higher suppression factors improve the possible measurements. For a detector with a resolution $\sigma_E\approx1$ eV, the activity for \Pu needs to be at least 10 times as high as for \T to reach the same sensitivity if the background suppression factor is 100. For suppression factor 10, the activity of \Pu needs to be around 50 times as high as for \T.

\paragraph{Data mangagement}
Additonal plots and the source code for this analysis can be found at: \href{https://github.com/AnaamMustafa/Pu241-Neutrino-Mass-Supplement}{Pu241-Neutrino-Mass-Supplement}

\section*{Acknowledgements}

\paragraph{Author contributions}
All authors contributed in equal amount to the analysis and publication.

\paragraph{Funding information}
This work was supported by Dutch Research Council (NWO) research grant:NWA.1292.19.231.

\clearpage
\bibliography{Pu241}

\begin{thebibliography}{10}
\providecommand{\url}[1]{\texttt{#1}}
\providecommand{\urlprefix}{URL }
\expandafter\ifx\csname urlstyle\endcsname\relax
  \providecommand{\doi}[1]{doi:\discretionary{}{}{}#1}\else
  \providecommand{\doi}{doi:\discretionary{}{}{}\begingroup \urlstyle{rm}\Url}\fi
\providecommand{\eprint}[2][]{\url{#2}}

\bibitem{katrin2025}
M.~Aker, D.~Batzler, A.~Beglarian, J.~Behrens, J.~Beisenk{\"o}tter, M.~Biassoni, B.~Bieringer, Y.~Biondi, F.~Block, S.~Bobien, M.~B{\"o}ttcher, B.~Bornschein \emph{et~al.},
\newblock \emph{Direct neutrino-mass measurement based on 259 days of {KATRIN} data},
\newblock Science \textbf{388}(6743), 180 (2025),
\newblock \doi{10.1126/science.adq9592},
\newblock \eprint{2406.13516}.

\bibitem{Project8-2017}
A.~A. Esfahani, D.~M. Asner, S.~B{\"o}ser, R.~Cervantes, C.~Claessens, L.~de~Viveiros, P.~J. Doe, S.~Doeleman, J.~L. Fernandes, M.~Fertl, E.~C. Finn, J.~A. Formaggio \emph{et~al.},
\newblock \emph{Determining the neutrino mass with cyclotron radiation emission spectroscopy - project 8},
\newblock Journal of Physics G: Nuclear and Particle Physics \textbf{44}(5), 054004 (2017),
\newblock \doi{10.1088/1361-6471/aa5b4f}.

\bibitem{qntm}
A.~A.~S. Amad, F.~F. Deppisch, M.~Fleck, J.~Gallop, T.~Goffrey, L.~Hao, N.~Higginbotham, S.~D. Hogan, S.~B. Jones, L.~Li, N.~McConkey, V.~Monachello \emph{et~al.},
\newblock \emph{Determining absolute neutrino mass using quantum technologies},
\newblock New Journal of Physics \textbf{27}(10), 105006 (2025),
\newblock \doi{10.1088/1367-2630/adc624}.

\bibitem{Ptolemy2019}
M.~Betti, M.~Biasotti, A.~Bosc\'a, F.~Calle, N.~Canci, G.~Cavoto, C.~Chang, A.~Cocco, A.~Colijn, J.~Conrad, N.~D'Ambrosio, N.~de~Groot \emph{et~al.},
\newblock \emph{Neutrino physics with the ptolemy project: active neutrino properties and the light sterile case},
\newblock Journal of Cosmology and Astroparticle Physics \textbf{2019}(07), 047 (2019),
\newblock \doi{10.1088/1475-7516/2019/07/047}.

\bibitem{Cheipesh2021}
Y.~Cheipesh, V.~Cheianov and A.~Boyarsky,
\newblock \emph{Navigating the pitfalls of relic neutrino detection},
\newblock Phys. Rev. D \textbf{104}, 116004 (2021),
\newblock \doi{10.1103/PhysRevD.104.116004},
\newblock ArXiv:2101.10069 [hep-ph].

\bibitem{Tozzini26}
A.~Casale, A.~Esposito, G.~Menichetti and V.~Tozzini,
\newblock \emph{$\ensuremath{\beta}$-decay spectrum of tritiated graphene: Combining nuclear quantum mechanics with density functional theory},
\newblock Phys. Rev. C \textbf{113}, 054607 (2026),
\newblock \doi{10.1103/gr8x-lf9f}.

\bibitem{ndg-pu241}
N.~de~Groot,
\newblock \emph{Plutonium-241 as a possible isotope for neutrino mass measurement and capture},
\newblock Journal of Physics G: Nuclear and Particle Physics \textbf{50}(5), 055106 (2023),
\newblock \doi{10.1088/1361-6471/acc5fc}.

\bibitem{audi1995}
G.~Audi and A.~H. Wapstra,
\newblock \emph{The 1995 update to the atomic mass evaluation},
\newblock Nuclear Physics A \textbf{595}(4), 409 (1995),
\newblock \doi{10.1016/0375-9474(95)00445-9}.

\bibitem{DEBIEVRE1983111}
P.~{De Bi{\`e}vre}, M.~Gallet and R.~Werz,
\newblock \emph{Determination of the 241pu half-life by isotope mass spectrometry},
\newblock International Journal of Mass Spectrometry and Ion Physics \textbf{51}(1), 111 (1983),
\newblock \doi{https://doi.org/10.1016/0020-7381(83)85032-3}.

\bibitem{np237data}
M.-M. B{\'e}, V.~Chist{\'e}, C.~Dulieu, X.~Mougeot, E.~Browne, V.~Chechev, N.~Kuzmenko, F.~Kondev, A.~Luca, M.~Gal{\'a}n, A.~L. Nichols, A.~Arinc \emph{et~al.},
\newblock \emph{Table of Radionuclides (Vol. 5 -- A = 22 to 244)},
\newblock Monographie BIPM-5. Bureau International des Poids et Mesures, S{\`e}vres, France (2010).

\bibitem{Mougeot-betashape}
X.~Mougeot,
\newblock \emph{Betashape: A new code for improved analytical calculations of beta spectra},
\newblock EPJ Web of Conferences \textbf{146}, 12015 (2017),
\newblock \doi{10.1051/epjconf/201714612015}.

\end{thebibliography}

\end{document}